\documentclass{aastex}

\begin {document}

\title{The Salpeter plasma correction for solar fusion reactions}

\author{John N. Bahcall}

\affil{Institute for Advanced Study
	Princeton, New Jersey 08540}

\author {Lowell S. Brown}

\affil{Department of Physics, University of Washington
        Seattle, Washington 98195}

\author{Andrei Gruzinov}

\affil{Institute for Advanced Study
	Princeton, New Jersey 08540}

\author{R. F. Sawyer}

\affil{Department of Physics, University of California at Santa
Barbara,
	Santa Barbara, California 93106}

\begin{abstract}
    
We review five different derivations that demonstrate that the
Salpeter formula for the plasma corrections to fusion rates is valid
at the center of the sun with insignificant errors ($\sim$ percent).
We point out errors in several recent papers that have obtained a
variety of answers, some even with the wrong sign or the wrong
functional dependence.

\end {abstract}

\section{Introduction}
\label{sec:itroduction}

The plasma in the core of the sun is sufficiently dense that non-ideal
 gas corrections to nuclear reaction rates are significant.  The
 plasma coupling parameter is $ g = (e^2 / D ) (1 / kT ) $, where $D$
 is the Debye length and $T$ the temperature of the plasma. This
 parameter is the ratio of the Coulomb potential energy for two
 particles a Debye length apart to the kinetic energy in the
 plasma. Near the center of the sun, $ g \simeq 0.04 $.  

Recently, there have been a number of 
papers (Carraro,
Sch\"afer, \& Koonin 1988; Shaviv \& Shaviv 1996, 1997, 2000;
Savchenko 1999; Tsytovich 2000; Opher \&
Opher 2000a, 2000b; Lavagno \& Quarati 2000)\nocite{SS1,SS2,SS3,tsyt,opher1,opher2,LQ,sav,CSK}
 suggesting that the standard screening corrections, originally
 derived by Salpeter (1954)\nocite{sal}, need to be replaced by some other
 plasma physics correction, and that moreover the changes could lead
 to substantial improvements in the standard solar model and the
 predicted solar neutrino fluxes.

The motivation for many of these papers is to `solve the solar
neutrino problem' without invoking new weak interaction physics, such
as neutrino oscillations. However, the results of solar neutrino
experiments cannot be accounted for in this manner even if one goes to
the extreme limit of treating the nuclear reaction rates as free
parameters (see, e.g., Bahcall, Krastev \& Smirnov 1998; Hata,
Bludman, \& Langacker 1994; and Heeger \& Robertson 1996)
\nocite{bks,hata,robertson}. Some distortion of the energy spectrum of electron
type neutrinos is required.

The purpose of this paper is to highlight the compelling evidence for
the Salpeter screening formula under the conditions that are relevant
at the center of the sun, i.e., in the limit of weak screening. Our
goal is to show that a necessary (but not sufficient) condition for
the validity of a screening calculation is that the calculation must
yield the Salpeter result in the limit when $g $ is very small. We
also point out errors in some of the recent treatments of
screening. The raison d'etre for our paper is the requests that we
have had from colleagues for a written response to the numerous papers
claiming large new effects (all different) in the calculation of solar
fusion rates.

We summarize in \S~\ref{sec:independent} the results of five
different derivations that all yield the Salpeter formula for
screening. In \S~\ref{sec:incorrect}, we describe briefly the flaws
that lead to five different, non-Salpeter screening formulae. We
summarize our principal conclusions in \S~\ref{sec:summary}.

\section{Five  derivations that yield the Salpeter formula}
\label{sec:independent}

\subsection{Salpeter electrostatic derivation}
\label{subsec:electrostatic}

As shown by Salpeter (1954)\nocite{sal}, fusion rates are enhanced by
electrostatic screening. Here is the physical plausibility argument
used by Salpeter.

If one of the fusing ions has charge $Z_1e$, it creates an
electrostatic potential $\phi =(Z_1e/r)\exp (-r/D)$, where $r$ is the
distance from the ion, and $D$ is the Debye radius. For $r\ll D$,
$\phi =Z_1e/r-Z_1e/D$ is the Coulomb potential minus a constant
potential drop. This potential drop increases the concentration of
ions $Z_2$ in the neighborhood of $Z_1$ by the Boltzmann factor
[$\propto \exp(-Z_2 e \phi /kT)$, where $T$ is the plasma temperature]
\begin{equation}
\label{eqn:salpeter}
f_0~=~ \exp \left({ {Z_1Z_2e^2}\over {kTD} }\right) ~=~\exp
\left(Z_1Z_2g\right),
\end{equation}
Equation~(\ref{eqn:salpeter}) is
the Salpeter formula. According to Salpeter, the quantity $f_0$ is
the ratio of the true reaction rate
to the reaction rate calculated using the ideal gas formula.

Salpeter's derivation makes physically clear that electrostatic
screening causes an enhancement in the  density of fusing partners
by lowering the potential in the vicinity of a fusing ion. We shall
come back to this physical argument in \S~\ref{subsec:tsytovich}.

\subsection{WKB derivation}
\label{subsec:wkb}

The correction due to screening can be derived by calculating the
barrier penetration factor in the presence of a plasma.  Bahcall,
Chen, \& Kamionkowski (1998)\nocite{BCK} evaluated 
the barrier penetration for a
Debye-Huckel plasma and showed that the usual Gamow penetration
factor, $e^{-2\pi\eta}$,
is replaced by 

\begin{equation}
\label{eqn:average}
     \Gamma(E)~=~e^{-2\pi\eta} e^{x\pi\eta},
\end{equation}
where $x=x(E)= R_c/D$.  Here, $D$ is again the Debye-Huckel radius and 
$R_c$ is the classical
turning-point radius defined by $V_{\rm sc}(R_c)=E$. Averaging $\Gamma(E)$
over a Maxwell-Boltzman distribution, the effect of $e^{x\pi\eta}$ is just
to multiply the total reaction rate by the Salpeter correction, $f_0$.

This derivation is more rigorous than the Salpeter formulation, but is
perhaps less transparent physically. The WKB derivation shows that the
Salpeter formula is valid in the moderate density limit in which
Debye-Huckel screening is a good approximation to the charge density
distribution.

\subsection{Density matrix derivation}
\label{subsec:densitymatrix}

Gruzinov and Bahcall (1998)\nocite{GB} calculated the electron density in the
vicinity of fusing nuclei using the partial differential equation for
the density matrix that is derived in quantum statistical mechanics.
This is the first calculation to describe properly the electron
density close to the fusing nuclei. Given the electron density,
Gruzinov and Bahcall then evaluated screening corrections in a mean
field approximation by solving numerically the Poisson-Boltzman equation
for a mixture of electrons and ions. The electron density distribution
obtained from the density matrix calculation was included
self-consistently and iteratively in the mean field equation.

The mean-field calculation yields exactly the Salpeter result, $f_0$,
in the limit of low density. Higher order screening 
 corrections were evaluated and found to be of order $1$\% for all of
the important solar fusion reactions.

\subsection{Free-energy calculation}
\label{subsec:classicalfree}

Dewitt, Graboske, \& Cooper (1973)\nocite{DGC} gave 
a rigorous derivation of the fusion rate corrections in the week
screening limit based on the free energy of fusing ions. Stimulated by
one of the incorrect derivations of screening corrections, Bruggen \&
Gough (1997)\nocite{BG} 
explained why the free energy is useful in this context.

For a given relative position of the two ions, one considers the
electrostatic contribution to free energy from the rest of the
plasma. In fact, it is sufficient to calculate the free energy
correction for a single charge $Z$, $\delta F(Z)$. Then the rate
enhancement factor is
\begin{equation}
\label{eqn:free}
f_0=\exp \left( {-\delta F(Z_1+Z_2)+\delta F(Z_1)+\delta F(Z_2)\over T}\right) 
.
\end{equation}
In the limit of small plasma density -- the week screening limit --
this free energy correction can be calculated exactly. The result is
the Salpeter formula. 

Equation~(\ref{eqn:free}) has the physically obvious characteristics
that the enhancement is symmetric in $Z_1$ and $Z_2$ and goes to zero
in the limit of $Z_1$ or $Z_2$ going to zero.  We shall come back to
these physically obvious characteristics in
\S~\ref{subsec:tsytovich}.

\subsection{Quantum field theory derivation}
\label{subsec:quantumfield}

Brown \& Sawyer (1997)\nocite{BS} have developed a rigorous, general
 formulation for calculating the rate of fusion reactions in
 plasmas. The Brown-Sawyer formalism can be used to develop an
 unambiguous perturbation expansion in the plasma coupling parameter
 $g = e^2/DkT $. The general formula derived 
in Brown \& Sawyer (1997)\nocite{BS} reduces to
 the Salpeter correction to first order in $g$; this correction should
 suffice for solar model calculations.

The Salpeter correction is formally of order $e^3$.  Terms of order
 $e^4$ are also examined in Brown \& Sawyer (1997)\nocite{BS}. 
The leading correction
 comes from the fact that the electrons are slightly degenerate, so
 that the first-order effects of Fermi statistics must be
 retained. This small effect is explicitly computed in
 Brown \& Sawyer (1997)\nocite{BS}. The remaining terms 
of the formal $e^4$ order are of
 relative order $\ell /D$ to the basic Salpeter correction,
 where $\ell$ is either a ionic thermal wavelength $\lambda$ or the
 Coulomb classical turning point $r$ of the fusing ionic motion. An
 upper bound shows that these are negligible contributions.  These
 higher-order calculations provide evidence, which goes beyond the 
 qualitative statement that the plasma is ``weakly coupled," that the
Brown-Sawyer  perturbation expansion is applicable in the solar domain.

\section{Papers claiming that Salpeter formula does not work}
\label{sec:incorrect}

\subsection{Dynamic screening}
\label{subsec:dynamic}

``Dynamic screening" (Carraro et al. 1988; see also Shaviv \& Shaviv
1997, 2000)\nocite{CSK,SS1,SS2}
is a generic name for attempts to rectify the following perceived
defect in the Salpeter argument: Approximately one-half of the squared
Debye wave number $D^{-2}$ comes from screening by electrons in
the plasma and one-half from ions.  As the two nuclei that are to fuse
approach each other, the electron speeds are fast enough for the
electronic cloud to adjust to the positions of the nuclei. But the
by-standing ions in the plasma may be thought to have a problem making
the adjustment, since their speeds are of order of the speeds of the
ions that are to fuse. Thus it could appear that the effects of the
ionic screening could be less than those in the Salpeter result, with
a consequent reduction in fusion rates.

However, Gruzinov (1998, 1999)\nocite{gruz1,gruz2} 
gave a general argument showing that in an
equilibrium plasma there is no such reduction. For the Gibbs
distribution
[probability of a state being occupied proportional to $\exp(-{\rm
energy}/kT)]$,
momentum and configuration probabilities are independent, and
velocities of fusing particles have no effect on screening. 

Moreover, in the most straightforward model for ``dynamical" screening
one can see how the ``dynamical" part of the correction terms get
canceled. The paper (Carraro et al. 1988)\nocite{CSK} calculates a modified potential
function for the fusing ions by following the motion of test bodies
with positive charge approaching one another in a plasma characterized
by the standard dynamic plasma dielectric constant. This modified
potential produces corrections in fusion rates that are perturbatively
of order $e^3$ times the uncorrected rates, as are the Salpeter
effects when expanded. However, in 
appendix D of Brown \& Sawyer (1997)\nocite{BS} it is
shown that resulting modifications of the Salpeter result are exactly
cancelled (to order $e^3$) when processes are included in which the
plasma has been excited or de-excited in a Coulomb interaction with
one of the incoming ions. This is the reason that a calculation based
only on a modified potential for elastic scattering fails.

To summarize, ``dynamical screening" results, both in their simple
realization in Carraro et al. (1988)\nocite{CSK}, and in 
any calculation that implements
the qualitative argument given above, are refuted both by the
general argument regarding the factorability of the distribution
function  and by explicit calculation.

\subsection{Unconventional interpretation of the Gibbs distribution}
\label{subsec:unconventialgibbs}

Opher \& Opher (2000b)\nocite{opher2} propose, in order 
to support their version
of dynamic screening, a different interpretation of the Gibbs
probability distribution than is given in any of the standard
treatments of statistical mechanics.  Opher and Opher claim that the
Gibbs distribution cannot be decomposed into uncoupled position and
velocity factors; they assert that the momentum variables must be
regarded as position dependent. This assertion confuses the concept of
a trajectory with a probability distribution.

The claim by Opher and Opher contradicts one of the foundations of
 statistical mechanics, as can be seen by consulting any of the
 standard expositions of the subject. For example, Landau \& 
 Lifchitz (1996)\nocite{landau96} stress: ``...the probabilities for momenta
 and coordinates are independent, in the sense that any particular
 values of the momenta do not influence the probabilities of the
 various values of the coordinates, and vice versa.''

The idea that coordinates and their conjugate momenta are independent
 statistical variables is familiar from elementary quantum mechanics
 where one calculates the phase space for a free particle as
 proportional to $d^3x d^3p$, the product of the differential volume
 in space and the differential volume in momentum.

\subsection{Cloud-cloud interaction}
\label{subsec:cloudcloud}

Shaviv and Shaviv (1996)\nocite{SS3} claimed that the screening energy,
${Z_1Z_2e^2}/D$, that appears in the exponent of Sapeter's formula
should be multiplied by a factor of $3/2$. They argued that a proper
inclusion of the electrostatic interaction between screening clouds
surrounding the fusing ions should lead to a modification of the
Salpter formula. As explained by Bruggen \& Gough (1997)\nocite{BG}, the
Shaviv and Shaviv treatment amounts to evaluating the potential
energy, $V$, for use in Schroodinger's equation by setting $V =
(\partial U/\partial r)_T$ rather than using the correct expression $V
=(\partial U/\partial r)_S$, where the subscripts $T$ and $S$ indicate
derivatives taken at constant temperature or constant entropy.

\subsection{Unconventional  statistics}
\label{subsec:statistics}

There are claims in the literature 
(see e.g., Savchenko 1999; Lavagno \& Quarati 2000)\nocite{LQ,sav} that the
usual Salpeter expression does not apply because standard statistical
mechanics (Gibbs distribution) is not valid; different statistical
distributions are proposed. 
There are at least three reasons why these
(and other authors) suppose that the Gibbs distribution is not valid
in the sun. These three reasons are: 1) perhaps there is not enough time
for statistical equilibrium to be established; 2) perhaps there are
interactions which distort the phase space distribution; and 3) perhaps the
Gibbs distribution is not the correct equilibrium distribution. We
discuss  these three possibilities in the following subsections.

\subsubsection{There is not enough time}
\label{subsubsec:notime}

Some of the suggested distributions seem to be based upon the
assumption that the core of the Sun is not in thermodynamic
equilibrium, and that there exist deviations from the Gibbs
distribution.  Both analytic calculations and Monte Carlo simulations
show that the energy distribution of ions in a plasma rapidly
approaches a Gibbs distribution on the time scale for the exchange of
a major fraction of the typical particle energy among the interacting
ions (see, e.g., MacDonald, Rosenbluth, \& Chuck
1957)\nocite{rosenbluth}.

There is a slight departure from statistical equilibrium in the energy
distribution of ions in the solar core, but the magnitude of the
effect is too small to be of significance for any measurable quantity.
The burning of nuclei in the sun is a non-equilibrium process, which
causes a departure from the ideal Gibbs distribution.  The magnitude
of the deviation, $\delta$, is of order the Coulomb collision time,
$\tau_{\rm Coulomb}$, over the nuclear burning time (Bahcall
1989)\nocite{bahc}. For the solar core,

\begin{eqnarray}
\delta  &~=~& { \tau_{\rm Coulomb} \over 
\tau_{\rm nuclear}}\nonumber\\
 &~=~&  10^{-28}\left[ \left({\tau_{\rm
nuclear} \over 10^{10} ~{\rm yr}}\right) \left({20 {\rm keV}\over E}\right)^{3/2}
\left({\rho\over 150 ~{\rm g ~cm}^{-3}}\right) \right]^{-1}.
\label{eqn:equilibrium}
\end{eqnarray}
The characteristic times for the most important solar fusion reates
range from $10^2$ yr to $10^{10}$ yr.
For purposes of calculating solar fusion rates, the solar interior is
in almost perfect thermodynamic equilibrium.

\subsubsection{Phase space distortion}
\label{subsubsec:phasespace}

The rate, R, for a binary nuclear reaction can be written symbolically
as

\begin{equation}
R ~\propto \int \!\! \int {d^3p_1 \, d^3p_2 \, \exp(-E/kT)\, |<f|H|i>|^2} .
\label{eqn:binaryrate}
\end{equation}
The term $d^3p_1d^3p_2$ in Eq.~(\ref{eqn:binaryrate}) represents the
free-particle density of states calculated when the particles are very far
separated; the Gibbs distribution is represented by the exponential;
and the interactions are described by the matrix element of the
Hamiltonian between initial and final states, $|<f|H|i>|^2$.
   
The basic error made by some authors (see e.g. Savchenko 1999)\nocite{sav} 
is to confuse the role
of the density of states, which can be calculated when the particles
are at very large separations ($d^3p_1d^3p_2$), with the role of the
interactions ($|<f|H|i>|^2$), which occur when the particles are very
close together.
 
\subsubsection{The Gibbs distribution is not the correct equilbrium distribution}
\label{subsubsec:notgibbs}

Many areas of modern physics, including large branches of condensed
matter physics, as well as many classical subjects are successfully
described by conventional statistical mechanics. There is no
convincing evidence for any phenomena that lie outside the domain of
standard statistical theory, which is described in the classical works
of, e.g., Tolman (1938)\nocite{tolman}, Feynman
(1972)\nocite{feynman}, 
and Landau \& Lifschitz (1996)\nocite{landau96}.

\subsection{Tsytovich Suppression}
\label{subsec:tsytovich}

Tsytovich (2000)\nocite{tsyt} suggests that screening leads to suppression
rather than enhancement of fusion rates. There are two ways to see
that the result given by Tsytovich (2000)\nocite{tsyt} is 
wrong, namely, the calculated
sign of the effect is incorrect and the functional dependence upon the
charges of the ions is incorrect.

First consider the sign of the effect, suppression rather than
enhancement of the reaction rate. The original Salpeter discussion,
summarized in \S~\ref{subsec:electrostatic}, showed that screening
enhances the reaction rate by lowering the potential in the vicinity
of the fusing ions.  This is the basic physical effect which must be
described by any correct theory of screening and which the elaborate
treatment of Tsytovich (2000)\nocite{tsyt} fails to recover. 

Since the treatment of Tsytovich is apparently very general, one may
also consider a limiting case in which very large impurities of
charges $\pm Q$ are introduced into a plasma undergoing $p-p$
fusion. The impurity charges are hypothesized to be so large that they
dominate over electrons and protons in the electrostatic
interactions. In these circumstances protons will preferentially clump
around negative charges $-Q$. Locally, the proton density will
increase and fusion will proceed faster. In this case, just as in the
general case discussed by Salpeter, electrostatic screening enhances
rather than suppresses fusion.

The result of Tsytovich (2000)\nocite{tsyt} does not pass 
an even more basic check. In
the week screening limit, the Salpeter
formula can be written as 
\begin{equation}
f_0{\rm (Salpeter)}~=~1~+~gZ_1Z_2
\label{eqn:salpeterlimit}
\end{equation}
 where $g$ does not depend on the charges of the reacting particles
$Z_1$, $Z_2$.  In the limit when one of the reacting particles has a
vanishingly small charge, the Salpeter screening effect goes to zero,
that is the screening enhancement $f_0=1$. The Tsytovich formula has a
different structure,

\begin{equation}
f{(\rm
Tsytovich)}~=~1~-~g_1Z_1^2 ~-~ g_2Z_2^2, 
\label{eqn:tsytovichlimit}
\end{equation}
which is a compact and revealing way to write Eq.~(11) 
of Tsytovich (2000)\nocite{tsyt}.  
Thus, $f{(\rm Tsytovich)}~\not=~ 1$  if one of the
particles is neutral, which is obviously incorrect.

\section{Summary and discussion}
\label{sec:summary}

There is only one right answer, but there are many wrong answers. 

We have reviewed five different derivations that all yield the
Salpeter screening formula, Eq.~\ref{eqn:salpeter}, in the weak screening
limit that is applicable to the solar interior. These derivations are:
the original Salpeter electrostatic argument, the WKB barrier
penetration calculation, the quantum statistical density matrix
evaluation, the free-energy calculation, and the rigorous quantum
field theory derivation.

In recent years, a number of authors have given alternative
expressions, each different from all the others, for the weak field
screening limit. We have described briefly in \S~\ref{sec:incorrect} 
the basic reason why each of these different non-Salpeter formulae are
incorrect. 

What can one  say about some future claim to have discovered an error in the
weak screening limit? Most readers, even those actively concerned with
fusion reactions in stellar interiors, do not have the time to examine
each of the claims for a new answer that are published. 

We suggest instead that the burden of proof should be upon the authors
claiming to have a result that differs from the Salpeter
formula. Discriminating readers may require of authors that claim to
have found a new answer that the authors first demonstrate fatal
errors in each of the five different derivations of the Salpeter formula that
are discussed in \S~\ref{sec:independent}.

\end{document}